\documentclass[iop,useAMS,usenatbib,apj,tighten]{emulateapj}
\usepackage{graphicx,amsmath,amssymb,subfigure,rotating,epsfig}

\def\simgt{\mathrel{\lower0.6ex\hbox{$\buildrel {\textstyle >} \over {\scriptstyle \sim}$}}}
\def\simlt{\mathrel{\lower0.6ex\hbox{$\buildrel {\textstyle <} \over {\scriptstyle \sim}$}}}

\newcommand{\Msolar}{\mbox{\,$\rm M_{\odot}$}}          

\newcommand{\mg}{Mg{\sc ii}\,\,}

\shorttitle{SERVS - Environments of High-$z$ QSOs}
\shortauthors{J.~T.~Falder et al.~2010}

\begin{document}

\title{The {\em{Spitzer}} Extragalactic Representative Volume Survey (SERVS):\\The
  Environments of High-$z$ SDSS Quasi-Stellar-Objects}

\author{J.~T.~Falder\altaffilmark{1},~J.~A.~Stevens\altaffilmark{1},~Matt~J.~Jarvis\altaffilmark{1},~D.~G.~Bonfield\altaffilmark{1},~M.~Lacy\altaffilmark{2},~D.~Farrah\altaffilmark{3},~S.~Oliver\altaffilmark{3},~J.~Surace\altaffilmark{4},\\~J.-C.~Mauduit\altaffilmark{4},~M.~Vaccari\altaffilmark{5},~L.~Marchetti\altaffilmark{5},~E.~Gonz\'{a}lez-Solares\altaffilmark{6},~J.~Afonso\altaffilmark{7,8},~A.~Cava\altaffilmark{9},~N.~Seymour\altaffilmark{10}}

\altaffiltext{1}{Centre for Astrophysics Research, Science \& Technology
  Research Institute, University of Hertfordshire, Hatfield, AL10 9AB, UK}
\altaffiltext{2}{NRAO, 520 Edgemont Road, Charlottesville, VA 22903, USA}
\altaffiltext{3}{Department of Physics and Astronomy, University of Sussex,
  Falmer, Brighton, BN1 9QH, UK} 
\altaffiltext{4}{Infrared Processing and
  Analysis Center/Spitzer Science Center, California Institute of Technology,
  Mail Code 220-6, Pasadena} 
\altaffiltext{5}{Department of Astronomy, Vicolo
  Osservatorio 3, University of Padova, I-35122, Padova, Italy}
\altaffiltext{6}{Institute of Astronomy, Madingley Road,
    Cambridge CB3 0HA, U.K.}
\altaffiltext{7}{Observat\'{o}rio Astron\'{o}mico de
  Lisboa, Faculdade de C\^{e}ncia, Universidade de Lisboa, Tapada da Ajuda,
  1349-018, Lisbon, Portugal}
\altaffiltext{8}{Centro de Astronomia da Univeridade de Lisboa, Lisbon,
       Portugal}
\altaffiltext{9}{Institutio de Astrof\'{i}sica de Canarias, C/V\'{i}a
  L\'{a}ctea s/n. 38200, La Laguna, Tenerife, Spain}
\altaffiltext{10}{Mullard Space Science Laboratory, UCL,
  Holmbury St Mary, Dorking, Surrey, RH5 6NT}

\email{J.T.Falder@herts.ac.uk}

\begin{abstract}
This paper presents a study of the environments of SDSS Quasi-Stellar-Objects (QSOs) in
the {\em Spitzer\/} Extragalactic Representative Volume Survey (SERVS). We
concentrate on the high-redshift QSOs as these have not been
studied in large numbers with data of this depth before. We use the
IRAC $3.6$-$4.5\mu$m colour of objects and ancillary $r$-band data to
filter out as much foreground contamination as possible. This technique allows
us to find a significant ($>4$-$\sigma$) over-density of galaxies around QSOs
in a redshift bin centred on $z\sim2.0$ and a ($>2$-$\sigma$) over-density of
galaxies around QSOs in a redshift bin centred on $z\sim3.3$. We compare our
findings to the predictions of a semi-analytic galaxy formation model, based on
the $\Lambda$CDM {\sc{millennium}} simulation, and find for both redshift
bins that the model predictions match well the source-density we have measured from
the SERVS data.
\end{abstract}

\slugcomment{A\sc{p}J, In Press}


\section{Introduction}

The {\em{Spitzer}} Extragalactic Representative Volume Survey (SERVS; Mauduit
et al. 2011, in prep) is a warm {\em Spitzer\/} survey at $3.6$ and $4.5 \mu$m
which will cover an area of 18 deg$^2$ in fields already extremely well-studied
and hence with a large amount of ancillary data. The survey reaches depths of
$\sim1\,\mu$Jy allowing $L_*$ galaxies to be observed out to $z\sim5$, thus
making it ideal for studying the environments in which AGN reside out to these
epochs.

It is now widely accepted that high-luminosity active galactic nuclei (AGN)
harbour accreting super-massive black holes with masses of the order $10^{7-9}
\Msolar$, implying that their host galaxies are amongst the most massive in
existence at their respective epochs. Indeed, many studies have now shown that
the most luminous types of AGN preferentially reside within fields containing
over-densities of galaxies (e.g. \citealt{Hall98,Best03,Wold03,Hutchings09}) as
would be expected of the most massive galaxies at any epoch. These points
support the idea that luminous AGN can be utilised as signposts to extreme
regions of the dark matter density and thus the most massive dark matter haloes
(e.g. \citealt{Pentericci00,Ivison00,Stevens03}) at high-redshift. Combining
this technique with large multiwavelength surveys, like the Sloan Digital Sky
Survey (SDSS; \citealt{Abazajian09}) which has identified more than 120000
broad-line quasi-stellar objects (QSOs) up to some of the highest measured
redshifts (i.e. $z$=6.4, \citealt{Fan03}; \citealt{Willott03}) has opened up a
new era in AGN research.

Recently the {\em Spitzer\/} Space telescope has been at the forefront of this
work due to its currently unique sensitivity to hot dust, which is an important
component of most AGN unified theories \citep{Antonucci93}. It has also been
utilised in the search for high-redshift galaxy clusters
(e.g.~\citealt{Eisenhardt08}, \citealt{Wilson09}, \citealt{Papovich10}). Its
wavelength range, particularly that of the Infrared Array Camera
(IRAC;~\citealt{Fazio04}), provides the necessary extension needed to take
cluster finding techniques to $z > 1$. Indeed it has led to the highest known
spectroscopically confirmed cluster to date (\citealt{Papovich10};
\citealt{Tanaka10}) at $z=1.62$, found solely from an over-density of IRAC
sources in the XMM-LSS field of the Spitzer Wide-Area Infrared Extragalactic
(SWIRE;~\citealt{Lonsdale03}) survey. Many of these methods make use of the
colour space that IRAC provides since it offers a useful way to select those
galaxies that are most likely to be at high-redshift. In addition, the negative
{\em{k}}-correction caused by the 1.6 $\mu$m peak in stellar emission moving
into the mid-infrared waveband means that IRAC can efficiently reach the depths
required to study the high-redshift universe in technically achievable exposure
times.

\begin{figure}
\centering
\includegraphics[width=0.95\columnwidth]{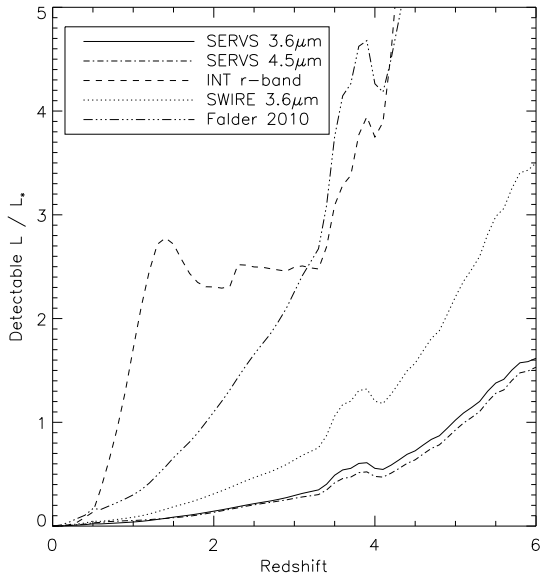}
\caption{The limiting magnitudes in terms of $L_*$ versus redshift for the INT
  $r$-band data (dashed line) as well as the SERVS $3.6 {\mu}m$ and $4.5
  {\mu}m$ data (solid and dash-dot lines respectively). Also shown for
  comparison is the SWIRE survey's $3.6 {\mu}m$ data (triple-dot-dash line) and
  the $3.6 {\mu}m$ data used in \citet{Falder10}. Full details of the models
  used to construct this figure are given in the text.}
\label{fig:L_star}
\end{figure}

These features that have proved so useful for cluster finding are also very
useful for the study of the environments of AGN at high-$z$. Recently in
\citet{Falder10} over-densities were found at $z\sim1$ in IRAC data around a
large sample of SDSS QSOs and radio galaxies; it was also found that radio-loud
AGN reside in, on average, denser environments. The depth of these observations
allowed for the detection of a $\sim 0.7L_*$ galaxy at the redshift of the AGN
($z\sim1$) and we could have detected an $L_*$ galaxy at $z\sim2$. On the other
hand, SERVS will allow the detection of $L_*$ galaxies at $z\sim5$, see
Fig.~\ref{fig:L_star}. The main difference in the sample used by
\citet{Falder10} and the sample used in this paper is that SERVS is a
blank-field rather than a targeted survey. This means the sample of AGN is
selected due to being in the survey region rather than based on other criteria
related to the AGN. Hence, we will not have a sufficient number of powerful
radio-loud AGN to look for similar effects as seen in \citet{Falder10} because
these are rare and require large areas or snapshot surveys to study in large
numbers. However, the almost unrivalled combination of depth and area that
SERVS provides will allow us to see if luminous AGN in general are found in
over-densities at $z >> 1$.

There is another link between the study of AGN environments and the study of
high-redshift clusters. Finding clusters beyond $z\sim1.5$ is challenging and
requires large, deep sky surveys; for example \citet{Papovich10} required all
of the SWIRE survey (49 sq degrees) to locate a $z=1.62$ cluster. However, AGN
are extremely luminous and therefore detectable out to $z>6$ with shallow wide
surveys like the SDSS or by their radio emission in large area radio
surveys. Many authors have therefore used these signposts for high-density
regions in the universe for follow-up and have located high-redshift clusters
and proto-clusters in observationally efficient ways (e.g.,
\citealt{Pentericci00}, \citealt{Stern03}, \citealt{Venemans07},
\citealt{Doherty10}, \citealt{Galametz10}). At $z>3$ only a handful of these
proto-clusters have been detected to date, mainly around individual
radio-galaxies (e.g., \citealt{Overzier06, Overzier08}). Recently in a study
similar to this work \citet{Hatch10} have studied the environments of
$z\sim2.4$ radio galaxies. In their work they find potential proto-clusters
around three of their six targets, with good evidence that the excess objects are
blue star-forming galaxies. Furthermore, studying the environments of high-redshift AGN may
provide important constraints on the level of positive \citep[e.g.][]{Elbaz09}
or negative \citep[e.g.][]{Rawlings&Jarvis04} AGN driven feedback on galaxies
in the immediate environment of the AGN.
 
In this paper we take advantage of the depth of SERVS to study the environments
in which high-redshift QSOs reside. The layout of the paper is as follows: in
Sections \ref{obs} and \ref{sample} we discuss the observations and the sample
of QSOs, in Section \ref{analysis} we describe our analysis, we present our
results in Section \ref{results} before comparing them to the previous
work at $z\sim1$ in Section \ref{z1} and to models in Section \ref{model}, we
then finish with a summary in Section \ref{summary}. Throughout the paper we
have assumed a flat cosmology with H$_0=72$ km\ s$^{-1}$\ Mpc$^{-1}$,
$\Omega_{\rm{m}}=0.3$ and $\Omega_{\Lambda}=0.7$. All magnitudes are quoted in
the AB system unless explicitly stated otherwise.

\begin{figure}
\centering \includegraphics[width=0.95\columnwidth]{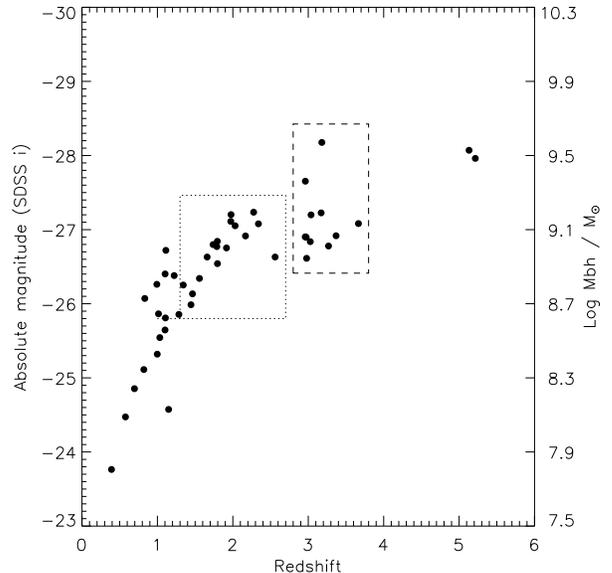}
\caption{Redshift vs rest frame optical absolute magnitude (SDSS $i$-band) for
  QSOs from the seventh data release of the SDSS quasar survey
  \citep{Schneider10} which are in the SERVS northern fields. The boxes show
  the two redshift bins we split the sample into for study as described in the
  text. On the right hand axis we show black-hole mass lower limits of these
  QSOs. These are calculated by assuming that the QSOs are accreating at the
  Eddington limit and using the relation from \citet{Rees84} with a bolometric
  correction factor of 15 \citep{Richards06a} to the SDSS $i$-band absolute
  magnitudes.}
\label{fig:L_z}
\end{figure}

\section{Observations and Source Extraction}

\label{obs}

The primary observations used in this paper are those from the {\em Spitzer\/}
Extragalactic Representative Volume Survey (SERVS). This is a warm
{\em Spitzer\/} survey using IRAC channels 1 and 2 ($3.6$
and $4.5$ ${\mu}m$ respectively). The data reach approximate 5-$\sigma$
depths of $\sim1~\mu$Jy (23.9 mag) at $3.6~{\mu}m$ and $\sim2~\mu$Jy (23.1
mag) at $4.5~{\mu}m$. Determining the depth of a large survey is non-trivial as
the coverage is not uniform in depth due to the overlaps of the scan
pattern. Small areas will therefore be deeper than the average, an important
consideration when measuring source density. We therefore need to cut the
catalogue at a flux level which ensures equal depth throughout the maps.

 Eventually SERVS will cover 18 deg$^2$ of the extremely well studied fields
 from the SWIRE survey. In this paper we make use of the SERVS overlap with the
 SDSS, thus restricting ourselves to the northern SERVS fields: Elais N1 (EN1)
 (1.01 deg$^2$ that overlaps the SDSS) and the Lockman Hole (4.93
 deg$^2$). Full details of the fields, observations and data reduction as well
 as the survey strategy will be given in Mauduit et al. (2011, in prep). In
 addition to the {\em Spitzer\/} data we make use of deep optical photometry
 from the INT (WFC) and KPNO (MOSAIC1), originally used by the INT WFS
 \citep{McMahon01} and the SWIRE optical imaging campaign \citep{Lonsdale03},
 but since expanded and re-reduced by Gonzalez-Solares et al. (2011, in
 prep). These data reach a 5-$\sigma$ depth of 24.2 mag in the $r$-band.

In Fig.~\ref{fig:L_star} we show the sensitivity of the SERVS data and the
$r$-band data in terms of $L_*$. For comparison we also show the $3.6~{\mu}m$
sensitivity from the SWIRE survey and that of the data used in
\citet{Falder10}. The main point to note is that while SWIRE detects $L_*$
galaxies at $z\sim2.5$, SERVS can detect them at $z\sim5$. This plot was
made using the restframe $K$-band luminosity function of \citet{Cir09} assuming
no evolution past $z=4$. All colour conversions between bands are derived using
a \citet{BC03} elliptical galaxy model with reddening of $A_v=0.8$ applied
according to the extinction law of \citet{Calzetti00} and the {\sc{hyperz}}
software package \citep{Bolzonella00}; this will be discussed at length in
Section.~\ref{colour_cut}. The $k$-correction is calculated using this model
from \citet{BC03} to place SEDs at various redshifts and comparing the
rest-frame and observed-frame flux in the $K$-band filter.

\begin{figure*}
\centering
\includegraphics[width=0.95\columnwidth]{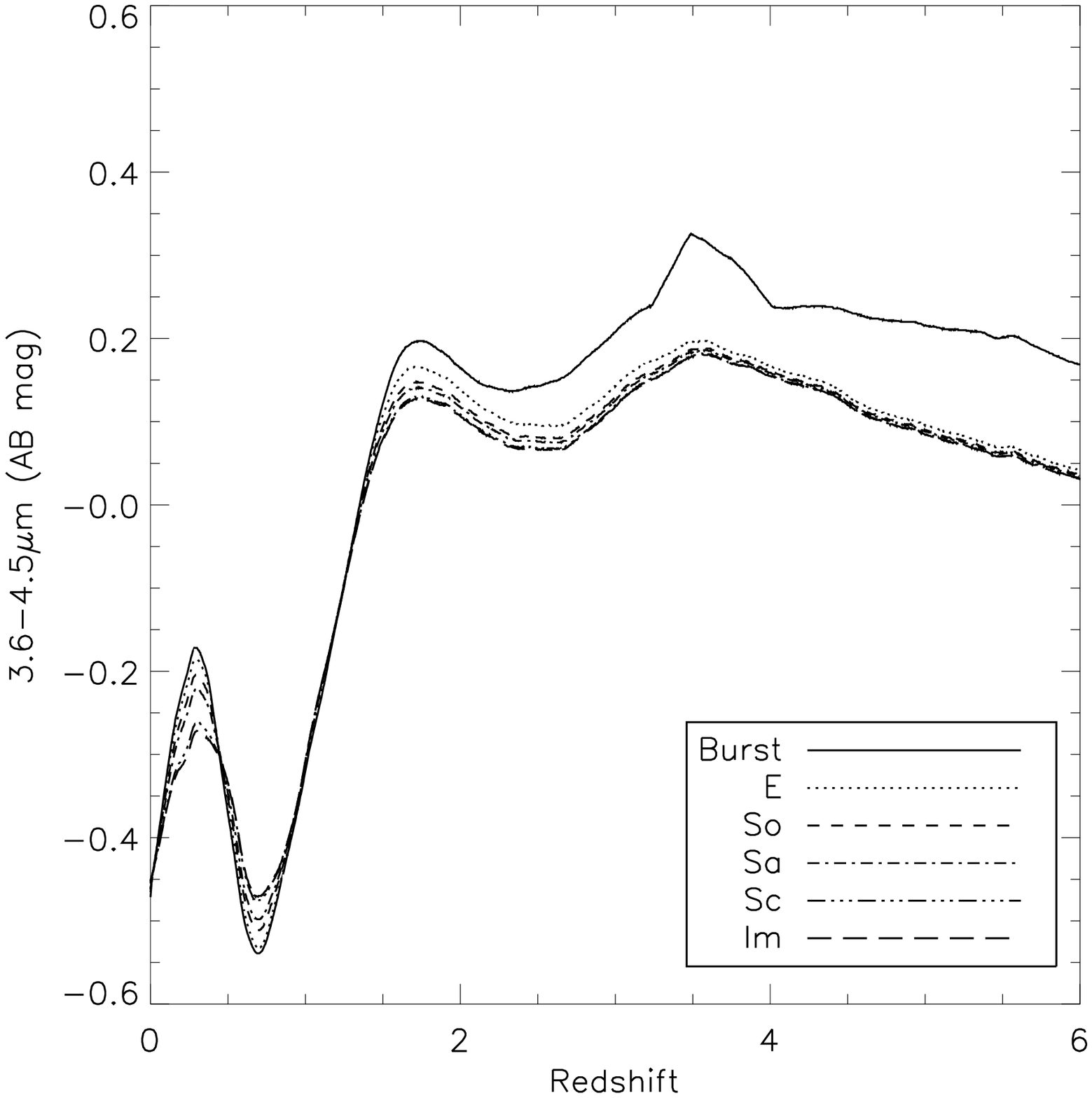}
\includegraphics[width=0.95\columnwidth]{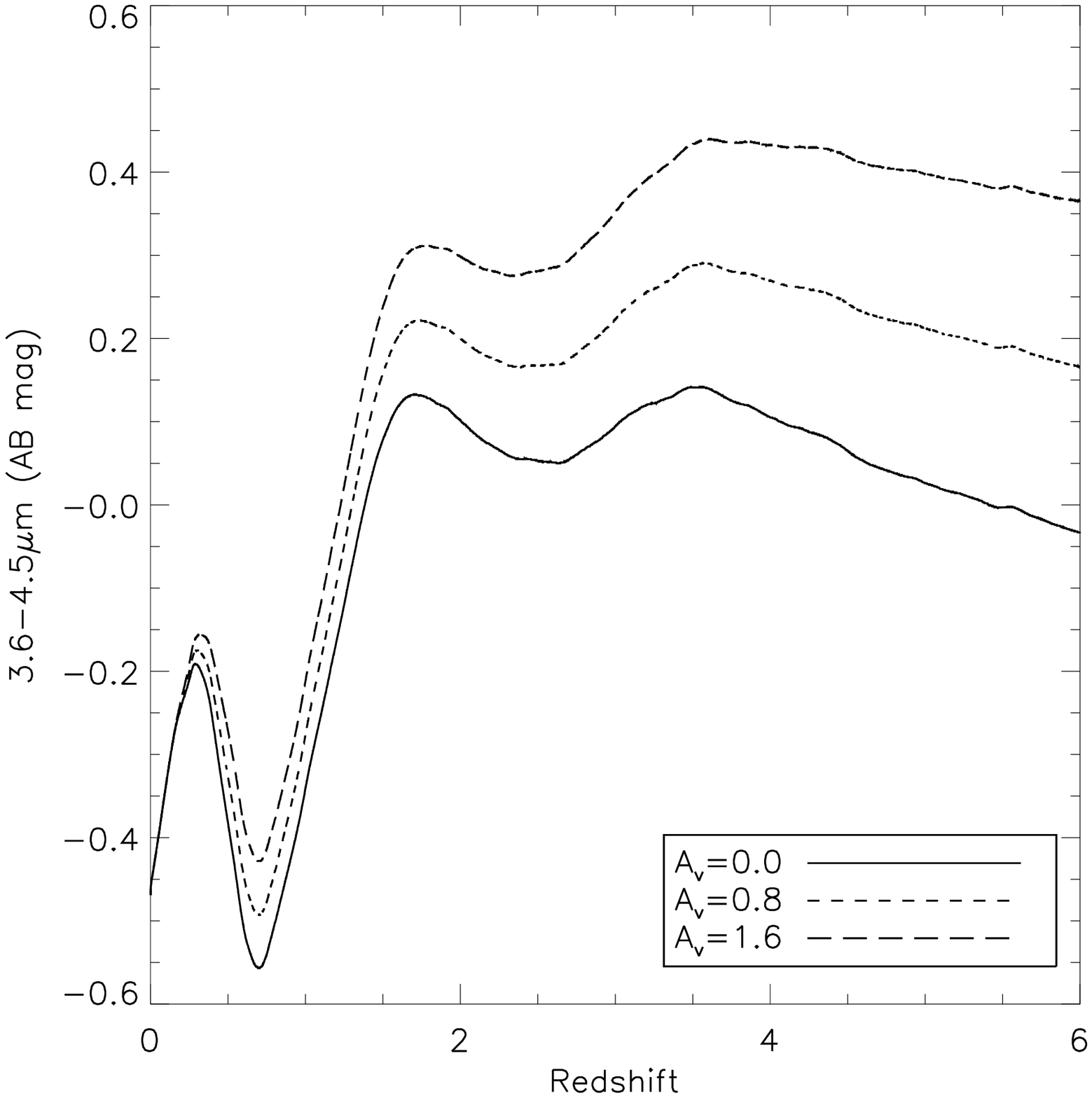}
\caption{IRAC $3.6$-$4.5\mu$m ($3.6 - 4.5 {\mu}m$) colour vs redshift, care of the
  \citet{BC03} stellar population models and the {\sc{hyperz}} software package
  \citep{Bolzonella00}. The left panel shows the range of colours produced with
  six different commonly used models. These consist of a single burst model,
  four exponentially declining SFR models representing elliptical, S0, Sa and
  Sc type galaxies with $\tau$ = 1,2,3 and 15 Gyr respectively, and a model
  with a constant SFR (Im). All models have a formation redshift of 10 and no
  reddening. In the right panel we show the elliptical model without reddening
  and with two models with $A_v$ = 0.8 and 1.6 applied according to
  \citet{Calzetti00}. This shows that a reasonable level of reddening is likely
  to have more of an effect on a galaxy's colour in this colour space than its
  star formation history.}
\label{fig:colour}
\end{figure*}

The catalogue we use is the SERVS data fusion catalogue (Vaccari et al. 2011, in
prep). This matches the single-band SERVS IRAC 3.6 and 4.5 $\mu$m catalogues
generated with the SExtractor software package \citep{Bertin&Arnouts96}, using
a search radius of 1.0 arcsec, computes an average coordinate (for sources
detected in both bands) and matches the resulting IRAC two-band catalogue with
ancillary photometric data-sets from the far-ultraviolet to far-infrared waveband
(e.g. GALEX, SDSS, CFHTCTIO/ESO/INT/KPNO/2MASS/UKIDSS/SWIRE) using a search radius
of 1.5 arcsec. Since we will use the $3.6$-$4.5\mu$m colour to select sources
likely to be at the correct redshift we therefore are, for the most part,
restricted to using only those sources which are detected in both bands.

\section{Sample}
\label{sample}

We identify QSOs in the SERVS regions by cross matching the SERVS source
catalogues with the seventh data release of the SDSS quasar survey
\citep{Schneider10} using the software package {\sc{topcat}} \citep{taylor05}
to select the SDSS QSOs in the overlap regions. In total we find 46 QSOs in the
SERVS northern fields; 5 in EN1 and 41 in the Lockman Hole. The small number in
EN1 is due to the SDSS only overlapping a small portion of the observed
region. These numbers include only QSOs that are at least $400$ arcsec from the
edges of the regions of equal coverage in both the ch1 and ch2 images. This
allows us to study the environments out to these distances without any effects
from the image edges. The distribution of the sample in the {\em{L-z}} plane is
shown in Fig. \ref{fig:L_z}. Six of the lower redshift QSOs are detected by the
FIRST radio survey \citep{Becker95} at 1.4GHz.

\section{Analysis}
\label{analysis}

\subsection{Radial search stacking}

To study the QSO environments we have employed the relatively simple technique
used in \cite{Best03} and \cite{Falder10}. This involves placing a series of
concentric annuli around each QSO and counting the number of sources that meet
our selection criteria, described in more detail later in Section
\ref{colour_cut}. We can then plot the source density as a function of radial
distance for each QSO. The annuli are kept to a fixed area as the radial
distance increases to keep the Poisson noise at a similar level from
bin-to-bin. The QSOs themselves are excluded from the search because including
them would bias the first annulus; this is done by not counting any sources
within $1$ arcsec of the QSO's SDSS coordinates. 

As we will be comparing our findings in this work to those from \cite{Falder10}
at $z\sim1$ we aim to conduct the analysis in a similar way. In that work, two
over-densities were reported, a sharp peak in the central source density within
$300$~kpc of the AGN and then a lower level over-density extending out to at
least $700$~kpc. This pattern has also been seen elsewhere in the literature,
for example in \citet{Best03} around powerful radio sources at $z\sim1.6$, and
by \citet{Serber06} for SDSS QSOs. When we look for over-densities within
$300$~kpc in this sample, while many AGN appear to have an over-dense first
annulus, they lack a significant detection. This is probably due to having far
fewer targets in the sample.  The sample used in \citet{Falder10} contained a
much larger number of AGN ($\sim$$170$) than are present in any of our redshift
bins or indeed the whole sample. \citet{Falder10} also found that it was not
possible to detect the $300$~kpc scale over-density with fewer than $\sim$$40$
randomly chosen AGN, suggesting that the central peak in source density is a
harder signal to detect than the larger scale over-density. 

To allow for easy comparison we use annular bins with a first bin physical
radius of $700$~kpc to match the largest search radius possible in the
\citet{Falder10} data, which was in turn fixed by IRAC's field of view. To take
into account the change in scale between different redshifts we adjust the
angular bin sizes for each QSO based on its redshift. This means the bins are
matched in terms of their physical size, where the radius of the first annulus is
$700$~kpc at the redshift of the QSOs.

To achieve a statistical result we will stack together the source density of
the QSOs in two coarse redshift bins using the raw number counts. To ensure we
are comparing like with like in the stacking analysis we match the range in
luminosity that we are sensitive to for each QSO. This is done in each redshift
bin by calculating what absolute magnitude the survey flux limit represents at
the highest redshift in that bin, and then adjusting the flux cut to ensure
this is matched for each of the other QSOs in that bin. As an upper limit we
work out what flux a $10L*$ galaxy would have at the maximum redshift of the
redshift bin and apply an upper cut on sources with fluxes greater than
this limit. The reason for choosing $10L*$ is to ensure we are not cutting any likely
associated galaxies, allowing for the significant uncertainty in the luminosity
function at these high redshifts.

The reasoning for adopting this method of analysis is that at these
high-redshifts other methods such as B$_{gq}$ and 2 point correlation functions
are difficult to calibrate correctly. The large error bars that result from
the assumptions made about the luminosity function and $k$-correction at these
redshifts make attaining a statistical result difficult.

\subsection{Galaxy colours}
\label{colour_cut}

\begin{figure}
\centering \includegraphics[width=0.95\columnwidth]{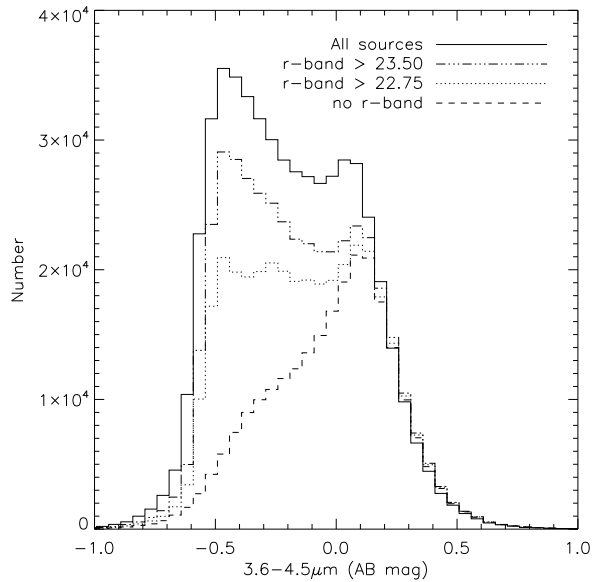}
\caption{Histogram of the IRAC $3.6$-$4.5\mu$m ($3.6 - 4.5 {\mu}m$) colour distribution
  in the SERVS source catalogues, shown for all sources (solid line), for
  sources with an $r$-band detection of 22.75 or fainter (dotted line), for
  sources with an $r$-band detection of 23.50 or fainter (dot and dashed line)
  and for those sources not detected in the $r$-band at all (dashed line). This
  figure clearly shows that most sources with $3.6$-$4.5\mu$m~$> 0.2$ are undetected in
  the $r$-band and therefore most likely to be at high-$z$, as
  Fig.~\ref{fig:colour} would suggest.}
\label{fig:colour_hist}
\end{figure}

In order to increase our sensitivity to galaxies at the same redshifts as the
QSOs we have made use of the IRAC $3.6$-$4.5\mu$m colour (i.e. ch1$-$ch2). To
help decide on the correct colour cuts we have made use of the {\sc{hyperz}}
software package \citep{Bolzonella00} and the stellar synthesis models of
\citet{BC03}. In the left panel of Fig. \ref{fig:colour} we show the colour of
6 commonly used models versus redshift. These are a single burst model, four
exponentially decreasing star formation rate (SFR) models with timescales $\tau$=1, 2, 3
and 15 Gyr designed to represent elliptical, S0, Sa and Sc type galaxies
respectively and a model with a constant SFR (Im). What is clear from
Fig. \ref{fig:colour} is that, with the exception of the burst model all the
models produce a very similar $3.6$-$4.5\mu$m colour; at most the burst model
differs by only 0.15 mags. In the right panel we show the effect that reddening
has on the $3.6$-$4.5\mu$m colour; this is shown for the elliptical model
without reddening along with two models with $A_v$=0.8 and 1.6, added according
to the \cite{Calzetti00} reddening law. This plot shows that adding a
reasonable amount of reddening can have a bigger effect on this colour space at
high-$z$ than the choice of star formation history. In all cases we have
assumed a formation redshift for the models of $z$=10. Changing this to $z$=100
made virtually no difference, and although using $z$=5 does make a difference,
at most it makes the colour bluer by $\sim$0.1 magnitudes at high-$z$. The key
thing to note is that for $z>1.3$ this colour space provides a good method for
selecting galaxies most likely to be at high-redshift.

Since our sample spans a range of $0.3 < z < 5.3$ we apply different colour
cuts to encompass different parts of the sample. Where we interpret our results
in terms of $L_*$ we use the elliptical model with $A_v=0.8$ (see
Fig.~\ref{fig:L_star}). However, for our colour cuts on the data we experiment
with colours that would fit any of the models. There are several factors to
consider here, firstly the real colours of galaxies will contain significant
scatter. We only show two parameters that can scatter the colours, SF
history and extinction, but in reality there will be more, not to mention the
intrinsic scatter from measurement errors. In \citet{Papovich08} the scatter of
the $3.6$-$4.5\mu$m colour was shown by over-plotting data with spectroscopic redshifts
on to the model predictions; this was possible for $z<3$ and it showed that
there was a minimum of 0.2 mags of scatter at all redshifts. An especially
problematic feature that \citet{Papovich08} reported was a population of
galaxies at $z\sim0.5$ with a significantly redder colour than could be
predicted by any model or with reddening. These galaxies will certainly
contaminate the colour space predicted to be occupied by high-$z$
galaxies. These issues create problems as encompassing a range of 0.4
magnitudes of colour space will mean we include a large foreground
contamination, potentially washing out the signal from the high-$z$ galaxies we
are interested in finding. It may prove correct that we are more sensitive to
galaxies at the redshifts of interest by using a narrower region of colour space
which, while losing some galaxies at the correct redshift,
means we remove more contamination. The distribution of IRAC $3.6$-$4.5\mu$m colour
from the SERVS source catalogues is shown in Fig. \ref{fig:colour_hist}. When
compared to Fig. \ref{fig:colour} the spread of colours is reassuringly
similar.

We therefore need a way of optimising our colour cut criterion for each
redshift range. Ideally this would be done using a spectroscopic sample of
galaxies which are in the SERVS fields, similar to that used by
\citet{Papovich08}. We could then use different criteria and see which values
return the most galaxies at the redshift of interest compared to other
redshifts. However, the number of spectroscopic redshifts available in the
SERVS fields, or in general at these high-redshifts, is insufficient for this
type of analysis. Instead, we conducted a Monte-Carlo simulation to adjust the
colour-cut criteria used around the QSOs. This method allows both the upper and lower
colour cut to be adjusted in steps within ranges determined from
Fig. \ref{fig:colour}. We then measured the source density for each Monte-Carlo
run and adopted the colour cut which gave us the largest over-density with
respect to the background. This works on the assumption that the signal will
peak when we include the most galaxies associated with the QSOs compared to
contaminating galaxies. To get an idea of the probability that these
over-densities are real, and not just noise spikes, we can then conduct the
same experiment many times around randomly chosen locations in the SERVS maps,
avoiding the locations of the QSOs in our sample.

The inability to effectively remove foreground contamination with this colour
space for $z < 1.3$ makes the sub-sample at these redshifts harder to study
with this method. At these low redshifts the available ancillary data are able
to provide a better means to study environments using photometric redshifts or
alternative colour cuts. It is for this reason that in this paper we avoid the
lowest redshift part of the sample. There are also far larger studies of such
objects' environments already in the literature (for example \citealt{Yee87},
\citealt{McLure01}, \citealt{Wold01}, \citealt{KauffmannHeckman&Best}, and with
{\em Spitzer\/} in \citealt{Falder10}) and so we feel a study with the small number 
we have in this sample would add little to the work already done at these relatively low 
redshifts. In contrast, the environments of high-$z$ QSOs have not been well studied in large
numbers or with data of this depth before.

\subsection{Ancillary data cuts}

In addition to the IRAC colour cut discussed in Section~\ref{colour_cut} we use
the $r$-band data from the INT (which reaches a 5-$\sigma$ limiting magnitude
of 24.2) to cut as much additional foreground contamination as possible. This
will hopefully allow us to tackle to some degree the $z\sim0.5$ galaxies
mentioned previously, that contaminate the colour space of higher-$z$
galaxies. The effect of cutting all sources with an $r$-band detection on the
$3.6$-$4.5\mu$m colour space is shown in Fig. \ref{fig:colour_hist}. The cut
has the clear effect of removing around two thirds of the sources with a
negative $3.6$-$4.5\mu$m colour, which is consistent with Fig. \ref{fig:colour}
suggesting that these sources have $z < 1.5$. The other interesting feature is
that very few sources with $3.6$-$4.5\mu$m~$> 0.2$ are detected in the
$r$-band, which is again consistent with Fig. \ref{fig:colour}, suggesting that
these sources are likely to lie at $z > 3$.

\begin{figure}
\centering \includegraphics[width=0.95\columnwidth]{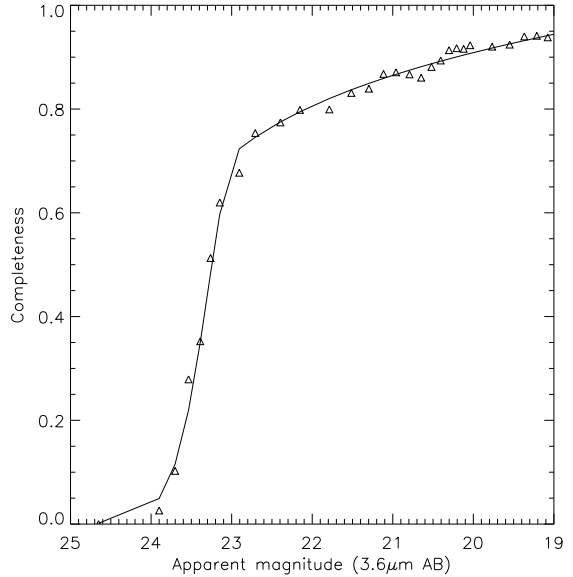}
\caption{The mean completeness fit for the IRAC ch1 data, taken as the average
  of the completeness in the outer few annuli, therefore away from the bright
  QSOs. The data points show the results from our extensive simulation and the
  curve is a combination of a fit to an empirical model of the form
  $Completeness=(S^a)/(b+cS^a)$ for completeness less than 0.7 and a power law
  for completeness greater than 0.7, described in more detail in the text.}
\label{fig:completeness}
\end{figure}

\begin{figure*}
\centering
\includegraphics[width=0.95\textwidth]{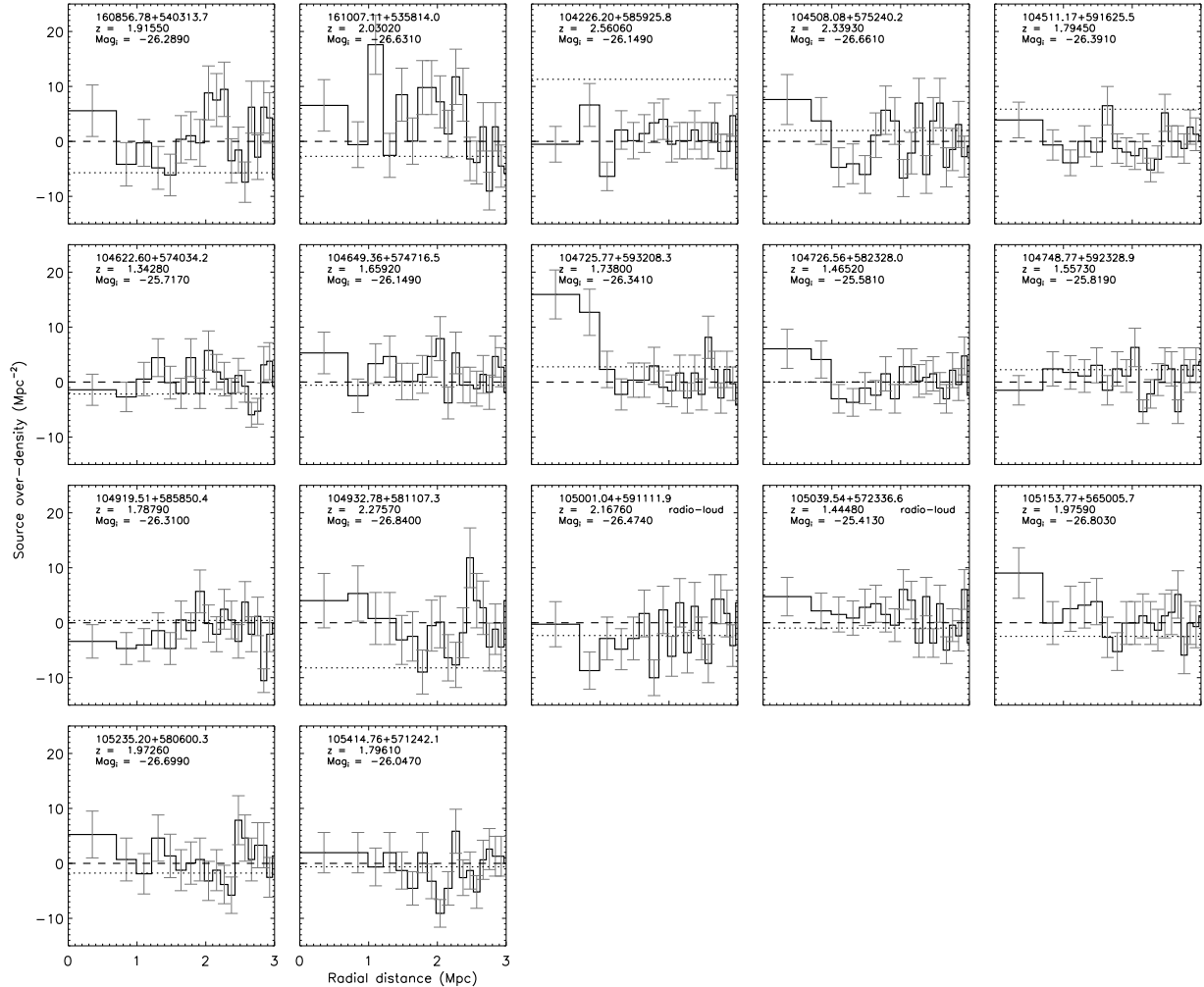}
\caption{Individual source over-density before being corrected for completeness
  vs radial distance for the 17 QSOs in the redshift range of $1.3 < z <
  2.7$. The first bin has a radius of $700$~kpc and the other bins are of the
  same area as the first. The error bars show the Poisson error on the number
  counts. The dashed line shows the subtracted local background level (zero
  level) determined from an annulus of $2$~Mpc$-400$ arcsec from the QSOs. The
  dotted line shows, for comparison, the global background as determined from
  taking the average source density in large apertures over the SERVS
  fields. Also labelled are the QSO's redshifts and absolute SDSS $i$-band
  magnitudes and those which are radio-loud.}
\label{fig:individual_1.3-2.7}
\end{figure*}

\subsection{Background level}

It is important to have a measure of the background level of sources expected
in the field with which to measure any over-densities against. There are
several ways in which this can be done, one would be to use a systematic offset
from the target and measure the source density in that region. This is known as
a blank or control field and provides an estimate of the local background
level. Another method when working with a large survey such as SERVS is to use
a global background where the average source density of a large area or the
whole survey is used. There are advantages and disadvantages to each method;
using the larger area washes out any fluctuations on small scales which might
affect a local value. However, if these local fluctuations are of an amplitude
which makes it important to take them into account in the background
determination then a local background will help if it is close enough to the
target. The trade off with this method is that being too close may result in
measuring the same structure of galaxies in both the background and target
field.

Both methods have been experimented with in this paper. The global background
has been calculated for each of the two SERVS fields, EN1 and the Lockman Hole,
by placing a series of $0.5^{\circ}$ radius circles on to the catalogues and
determining the average source density contained within them. The local values
have been calculated by using the average source density in an annulus which is
sufficiently far from the QSOs that it should not be measuring the same
structure. These annuli have a maximum radius of $400$ arcsec from the QSO to
ensure none of them fall off the image edges. The inner radius is determined
such that it should be 2~Mpc from the QSOs at their redshift. This should be
far enough that we are not likely to be measuring the same large-scale
structure that the QSOs reside in. Most evidence in the literature suggests
that, except for the largest galaxy groups and clusters, most of their members
are found within a radius $< 2$ Mpc (for example \citealt{Hansen05} and
\citealt{Papovich10}).

Obviously the background levels have to be determined in an identical way to
how we measure the source density in the environments of the QSOs and will be a
different value for each colour and $r$-band criterion that we use. The error
on the background is calculated as the Poisson error on the raw number counts
used for the background measurements. This is then scaled to the same area and
added in quadrature to the error on the source density measurements to get the
error on any over-density.

\subsection{Completeness}

In order to account for the incompleteness of the {\em Spitzer\/} data near the
flux limit of the survey we ran an extensive completeness simulation. This
largely followed the process used in \citet{Falder10} and full details are
given there. It involved cutting out regions of the {\em Spitzer\/} images
surrounding each QSO that measured $900 \times 900$ arcsec, i.e., 
large enough to include all annuli used in our analysis. We then inserted
10000 artificial sources into each of these cut outs for 40 different flux
levels. To avoid increasing source confusion these were added in batches of
1000 which meant that they were never too close together that they could become
confused with another artificial source. At each flux level we compared the
number of inserted sources to those found in the source catalogues. We consider
a source recovered if it is found within $1.2$ arcsec (2 pixels) of the
inserted location and within a factor of 2 of the inserted flux. The results of
this analysis are measured for each annulus separately which then enables us to
apply a completeness correction specific to each annulus. This means we account
for missed area in the vicinity of bright stars or indeed the QSOs. It was
shown in \citet{Falder10} that the bright QSO in the first annulus has the
effect of lowering the completeness in innermost annuli.

To eliminate the scatter in the measured completeness curves we then fitted
them with an empirical model of the form $Completeness=(S^a)/(b+cS^a)$
(\citealt{Cop06}) where $S$ is the 3.6 $\mu$m or 4.5 $\mu$m flux density and
$a$, $b$ and $c$ are constants that are fitted. It was found, however, that this
model alone was unable to provide a good fit at the knee between the bright end
and the steep slope; the data appear less complete here than would be
expected based on the S shape curve although this is at most a 10 per cent
effect. It is suspected this effect is likely due to source confusion as the SERVS
data are confused using the classical definition of 30 beams per source
(Mauduit et al. 2011, in prep). In order to overcome this effect we instead
fitted a power law to the data points when the completeness was greater than
0.70 (see Fig.\ref{fig:completeness}). The completeness is measured in both the
$3.6\mu$m and $4.5\mu$m images. We are then able for each detected source in
the real catalogues to calculate the correct completeness correction to apply
by multiplying the completeness fraction corresponding to the measured
$3.6\mu$m and $4.5\mu$m fluxes of the source.

We show the mean completeness curve for the $3.6\mu$m cutouts in
Fig.\ref{fig:completeness} based on the outer annuli, therefore away from the
bright QSOs. To make a conservative cut at the 50 per cent completeness value,
we use the flux density at which we are 50 per cent complete in
the first annulus. This gives a mean 50 per
cent completeness at $3.6\mu$m of $1.50\pm0.11\mu$Jy ($23.46^{+0.08}_{-0.07}$)
and at $4.5\mu$m of $1.77\pm0.08\mu$Jy ($23.28^{+0.05}_{-0.05}$). To be
conservative we only use sources detected with fluxes greater than the highest
flux level at which any of our QSO's first annuli were 50 per cent complete in
each channel; these values are $1.71\mu$Jy (23.32) at $3.6\mu$m and $1.89\mu$Jy
(23.21) at $4.5\mu$m.

\section{Results}
\label{results}

In order to have suitable numbers of objects in each redshift range we have
divided the sample into two redshift bins shown in Fig.~\ref{fig:L_z}. The
first is centred on $z\sim2.0$ and spans the range $1.3 < z < 2.7 $ and the
second is centred on $z\sim3.3$ and spans the range $2.8 < z < 3.8$. These bins
were chosen since there is a natural divide in the sample at $z\sim2.8$ which
gives samples of 11 and 17 QSOs in the two bins. This is a trade off between
having enough QSOs in the sub-samples whilst restricting the sample to as small
a fraction of cosmic time as is possible. As mentioned earlier for QSOs with
$z<1.3$ the colour space is contaminated by lower redshift galaxies and these
are already better studied in the literature. We also looked at the two
$z\sim5$ QSOs but failed to gain a significant detection of galaxies in their
environments, which is not surprising for such a small sample.

\subsection{$z\sim2.0$ sample}
\label{1.3 < z < 2.7  sample}

\begin{deluxetable}{cccc}  
\tabletypesize{\scriptsize}
\tablewidth{0.95\columnwidth}
\tablecaption{Table showing the set of upper and lower $3.6$-$4.5\mu$m colour
  criterion used for the Monte-Carlo analysis on the sample centred on
  $z\sim2.0$, along with each set's associated Poisson significance. Also
  shown is the percentage of times that this significance was achieved in the
  Monte-Carlo simulation around random points for this colour criterion. The
  colour criterion that gives the most significant over-density is shown in
  bold. The bottom two colour steps are designed for comparison to look above
  and below the region of colour space at which we expect to find an
  over-density, as would be expected these two criterion do not produce a
  significant over-density.\label{tab:lowz}}
\tablehead{
\colhead{Lower cut}          & \colhead{Upper cut}      &
\colhead{$\sigma_{Poisson}$}           & \colhead{Monte-Carlo \%}}
 \startdata
  -0.30  &   0.10  &   3.22  &   0.50  \\
  -0.30  &   0.15  &   4.01  &   0.10  \\
  -0.30  &   0.20  &   3.32  &   0.60  \\
  -0.30  &   0.25  &   3.13  &   1.00  \\
  -0.30  &   0.30  &   3.18  &   1.00  \\
  -0.25  &   0.10  &   3.29  &   0.40  \\
  {\bf -0.25}  &   {\bf 0.15}  &   {\bf 4.10}  &   {\bf 0.10}  \\
  -0.25  &   0.20  &   3.37  &   0.40  \\
  -0.25  &   0.25  &   3.17  &   0.80  \\
  -0.25  &   0.30  &   3.21  &   0.80  \\
  -0.20  &   0.10  &   3.23  &   0.40  \\
  -0.20  &   0.15  &   4.08  &   0.00  \\
  -0.20  &   0.20  &   3.29  &   0.50  \\
  -0.20  &   0.25  &   3.07  &   1.00  \\
  -0.20  &   0.30  &   3.12  &   1.20  \\
  -0.15  &   0.10  &   2.60  &   1.20  \\
  -0.15  &   0.15  &   3.56  &   0.20  \\
  -0.15  &   0.20  &   2.72  &   1.80  \\
  -0.15  &   0.25  &   2.50  &   3.10  \\
  -0.15  &   0.30  &   2.56  &   3.20  \\
  -0.10  &   0.10  &   2.03  &   3.80  \\
  -0.10  &   0.15  &   3.12  &   0.50  \\
  -0.10  &   0.20  &   2.23  &   4.30  \\
  -0.10  &   0.25  &   2.00  &   7.41  \\
  -0.10  &   0.30  &   2.08  &   7.01  \\
  -0.05  &   0.10  &   1.50  &  10.11  \\
  -0.05  &   0.15  &   2.76  &   1.00  \\
  -0.05  &   0.20  &   1.79  &   9.21  \\
  -0.05  &   0.25  &   1.56  &  13.71  \\
  -0.05  &   0.30  &   1.65  &  12.11  \\
 NA  &  -0.20  &   0.70  &  31.03  \\
   0.20  &  NA  &  -0.84  &  78.18  \\
\enddata
\end{deluxetable}

\begin{figure}
\centering
\includegraphics[width=0.95\columnwidth]{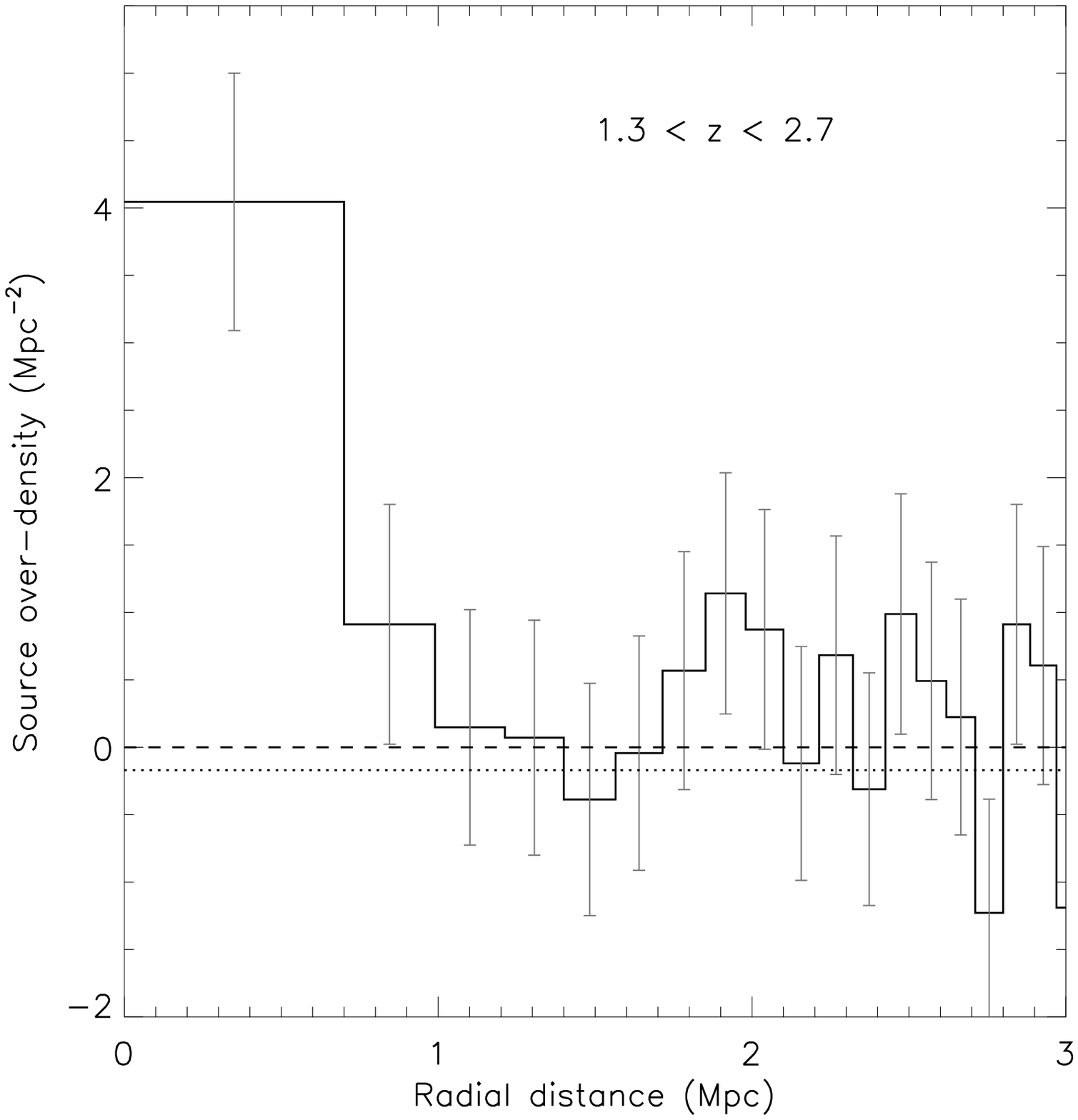}
\caption{Stacked source over-density before being corrected for completeness vs
  radial distance for the 17 QSOs in the redshift range of $1.3 < z < 2.7$. The
  first bin has a radius of $700$~kpc and the other bins are of the same area
  as the first. The error bars show the Poisson error on the number counts. The
  dashed line shows the subtracted local background level (zero level)
  determined from an annulus of $2$~Mpc$-400$ arcsec from the QSOs. The dotted
  line shows, for comparison, the global background as determined from taking the average source
  density in large apertures over the SERVS fields.}
\label{fig:1.3-2.7}
\end{figure}

\begin{figure}
\centering \includegraphics[width=0.95\columnwidth]{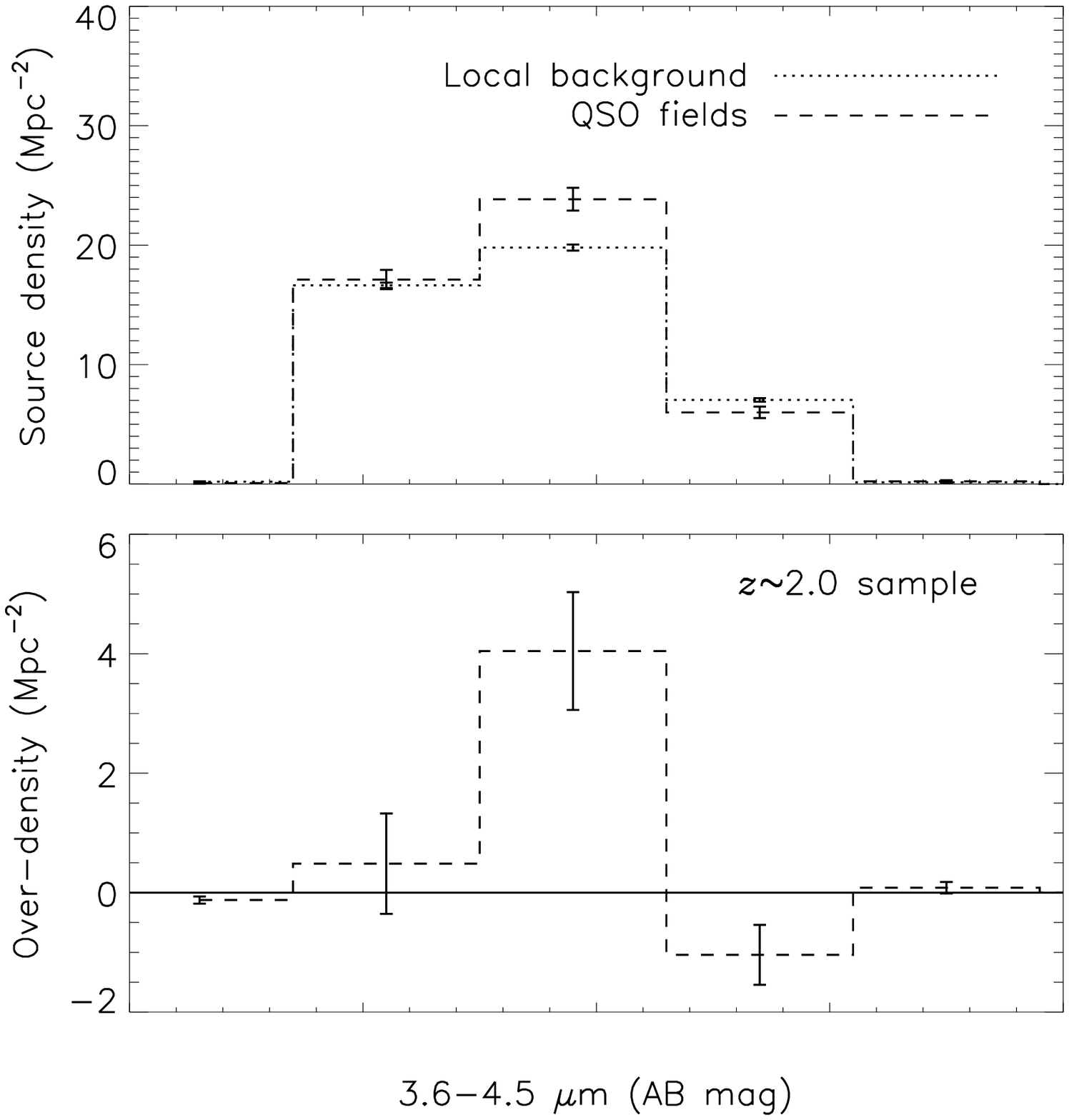}
\caption{Histograms showing the IRAC $3.6$-$4.5\mu$m colour of sources in the SERVS
  catalogues. In the top panel the dotted line shows the averaged local
  background source density surrounding the QSOs in the $z\sim2.0$ sample and
  the dashed line shows the averaged source density inside the first annuli
  surrounding these QSOs. The bottom panel shows the result of subtracting the
  local background from the source density in the first annulus.}
\label{fig:colour_histogram_z2}
\end{figure}

In the redshift range $1.3 < z < 2.7 $ there are 17 QSOs of which two are
detected by FIRST. These are 105001.04+591111.9 and 105039.54+572336.6 with
radio luminosities of Log$_{10}(\rm L_{1400}/
\rm{W\ Hz}^{-1}\ \rm{sr}^{-1})=25.39$ and $25.16$ respectively (calculated using the FIRST
radio flux and assuming a spectral index of 0.7). The flux limit at which we are
50 per cent complete at the maximum redshift of this part of the sample
corresponds to an absolute magnitude of -23.4 and so we restrict our search
around each QSO to galaxies brighter than this value (note that we ignore the
$k$-correction within the bin as this is at most a 0.02 magnitude effect). The adopted limit represents
galaxies which are roughly $0.7L_*$ or brighter in this redshift range.

 To choose a colour criterion for this sample we use Fig. \ref{fig:colour} to
 point us towards colours which may select galaxies at the redshift we are
 interested in. We then use our Monte-Carlo code to vary both the upper
 and lower colour cuts in steps within an appropriate range of model
 predictions. The results of this analysis are shown in Table \ref{tab:lowz} which shows
 that the most significant over-density of $4.10\sigma$ occurs in the stacked
 source density when a colour criterion of $-0.25 <$~$3.6$-$4.5\mu$m~$< 0.15$
 is used. In addition to the IRAC colour cut, we
 remove sources detected in the $r$-band with an apparent magnitude brighter
 than 23.5. This criterion corresponds to removing objects that are $\sim4L_*$
 or brighter according to our choice of models, meaning we are only excluding,
 if anything, the rarest galaxies associated with the QSOs.

The individual histograms of source over-density versus radial distance for
each of these 17 QSOs are shown in Fig.~\ref{fig:individual_1.3-2.7}; these
show the source density with the local background level subtracted. One of
these QSOs has a significant over-density around it at $>3\sigma$ level (given
by Poisson statistics), while several of the other first annuli are more than
1-$\sigma$ over-dense, which suggests stacking may produce a robust signal. It
is worth noting that neither of the two radio-loud QSOs (labelled) show any
sign of an over-density, which means we are confident that they will not bias
the stacked source-density as the results of \cite{Falder10} suggests they
might, probably due to their comparatively low radio luminosity.

The resulting stacked source density is shown in Fig.~\ref{fig:1.3-2.7}, which
shows an over-density within $700$\,kpc of the QSOs and is significantly above
the local background level at the 4.10-$\sigma$ (given by Poisson statistics)
level. The next annulus is also above the background level hinting that the
over-density extends to around the $1$~Mpc scale. If we exclude the QSO that
has an individually significant over-density this reduces to the 3.3-$\sigma$
level suggesting that the over-density we see in the stacked histogram is not
just around that one object. The global background while looking consistent in
most cases, seems to be too high or low in a few cases, suggesting they are in
a region with a locally high or low background density. Using the global
background has the effect of increasing the detected over-density to the
4.44-$\sigma$ level; we show where the global background level would be for
comparison with a dotted line in Figs.~\ref{fig:individual_1.3-2.7} and
\ref{fig:1.3-2.7}.

To put this choice of colour cut into context we show a histogram of the
$3.6$-$4.5\mu$m colour space in Fig.~\ref{fig:colour_histogram_z2} for both the
local background and the first annuli surrounding the QSOs; shown in the bottom
panel is a histogram of the result of subtracting the local background from the
first annulus. There is a clear over-density significant at the $4$-$\sigma$
level in the QSO fields, the location of which in this colour space is
consistent with it being in the redshift range of the QSOs.

When we run our Monte-Carlo code 1000 times on batches of 17 random locations
(to match the number of QSOs used) avoiding the QSO's locations in the process,
we find that we can generate similar sized over-densities only 0.1 per cent of
the time in the same colour space. This increases to only 0.5 per cent of the
time over all colour space sampled in this analysis (see Table
\ref{tab:lowz}). We are therefore confident that this over-density is real
and associated with the QSOs at the 99.5 per cent confidence level using this
method. The reason that this random field test does not generate Poisson
statistics is that in reality galaxies are clustered, and so the probability of
finding a second galaxy is not mutually exclusive of finding the first as is
the case for the Poisson distribution. It is worth noting that if we apply a
more simplistic colour cut $-0.25 <$~$3.6$-$4.5\mu$m, to remove only foreground
galaxies a 2.8-$\sigma$ (Poisson) over-density still remains.

Physically, the over-density in the first bin including a correction for
completeness corresponds to, on average, 7-10 brighter than $\sim0.7L_*$
galaxies, with our choice of models. This number is in excess of the local field level
around each QSO within $\sim700$~kpc, taking into account the range spanned by
the 1-$\sigma$ error bars in Fig.~\ref{fig:1.3-2.7}.

\subsection{$z\sim3.3$ sample}
\label{2.8 < z < 4.0  sample}
\begin{deluxetable}{cccccc}  
  \tabletypesize{\scriptsize} \tablewidth{0.95\columnwidth} \tablecaption{Table
    showing the same as in Table \ref{tab:lowz} but for the sample centred on
    $z\sim3.3$. This is shown both for the analysis down to the conservative 50
    per cent completeness limit (c50) and the 30 per cent completeness limit
    (c30). \label{tab:midz}} \tablehead{ \colhead{Lower cut} & \colhead{Upper
      cut} & \colhead{$\sigma_{c50}$} & \colhead{M-C \%$_{c50}$} &
    \colhead{$\sigma_{c30}$} & \colhead{M-C \%$_{c30}$}} \startdata
  -0.10  &   0.30  &   0.94  &  28.50  &   1.29  &  23.00  \\
  -0.10  &   0.35  &   1.33  &  19.90  &   1.77  &  14.70  \\
  -0.10  &   0.40  &   1.36  &  18.80  &   1.83  &  14.00  \\
  -0.10  &   0.45  &   1.63  &  13.90  &   2.08  &   9.10  \\
  -0.05  &   0.30  &   1.17  &  22.10  &   1.49  &  19.40  \\
  -0.05  &   0.35  &   1.57  &  14.70  &   1.99  &  11.70  \\
  -0.05  &   0.40  &   1.60  &  13.50  &   2.05  &  11.30  \\
  -0.05  &   0.45  &   1.88  &  10.00  &   2.31  &   7.60  \\
  0.00  &   0.30  &   0.79  &  31.90  &   1.08  &  26.20  \\
  0.00  &   0.35  &   1.24  &  20.70  &   1.64  &  16.10  \\
  0.00  &   0.40  &   1.27  &  19.20  &   1.70  &  15.80  \\
  0.00  &   0.45  &   1.58  &  14.10  &   1.98  &  10.90  \\
  0.05  &   0.30  &   1.61  &  10.80  &   1.77  &  11.50  \\
  0.05  &   0.35  &   2.04  &   6.90  &   2.33  &   5.40  \\
  0.05  &   0.40  &   2.06  &   7.10  &   2.38  &   5.40  \\
  0.05  &   0.45  &   2.37  &   4.20  &   2.66  &   3.50  \\
  0.10  &   0.30  &   1.18  &  19.40  &   1.36  &  17.60  \\
  0.10  &   0.35  &   1.69  &   9.60  &   2.01  &   7.30  \\
  0.10  &   0.40  &   1.71  &   9.90  &   2.06  &   7.50  \\
  0.10  &   0.45  &   2.06  &   6.00  &   2.38  &   4.30  \\
  0.15  &   0.30  &   1.59  &  10.20  &   1.95  &   5.70  \\
  0.15  &   0.35  &   2.14  &   4.40  &   2.62  &   2.20  \\
  0.15  &   0.40  &   2.14  &   4.40  &   2.65  &   2.60  \\
  0.15  &   0.45  &   2.51  &   1.70  &   2.99  &   0.70  \\
  0.20  &   0.30  &   1.60  &   9.20  &   2.11  &   2.80  \\
  0.20  &   0.35  &   2.22  &   2.90  &   2.86  &   0.70  \\
  0.20  &   0.40  &   2.19  &   2.90  &   2.85  &   0.50  \\
  {\bf 0.20}  &  {\bf 0.45}  &  {\bf 2.62}  &  {\bf 1.10}  &  {\bf 3.22}  &  {\bf 0.10}  \\
  0.45  &  NA  &   0.46  &   37.80  &   0.23  &  41.70  \\
  NA  &   0.05  &   0.55  &  36.10  &   0.75  &  31.90  \\
\enddata
\end{deluxetable}

In the bin that spans the redshift range $2.8 < z < 3.8$ there are 11 SDSS
QSOs. Using Fig. \ref{fig:colour} we again experiment, as in Section~\ref{1.3 <
  z < 2.7 sample} with our Monte-Carlo method of adjusting the colour
cuts. The results of this analysis are given in Table~\ref{tab:midz} which shows that
we find that a colour criterion of $0.20 <$~$3.6$-$4.5\mu$m~$< 0.45$ provides
the largest over-density. We again hope to minimise
contamination from the foreground by cutting all sources detected in the
$r$-band with an apparent magnitude brighter than 23.5.  In this redshift range
this criterion corresponds to cutting objects that are $\sim4L_*$ or brighter,
according to our choice of models, ensuring once again we are only excluding,
if any, the rarest galaxies associated with the QSOs. The flux at which we are
50 per cent complete corresponds to an absolute magnitude of -24.4 at the
maximum redshift of this samples range and so we restrict the search around
each QSO to galaxies brighter than this limit (we ignore the $k$-correction within the bin as
this is at most a 0.2 magnitude effect). Our adopted limit represents galaxies which are
roughly $1.1L_*$ or brighter in this redshift range.

\begin{figure*}
\centering
\includegraphics[width=0.95\textwidth]{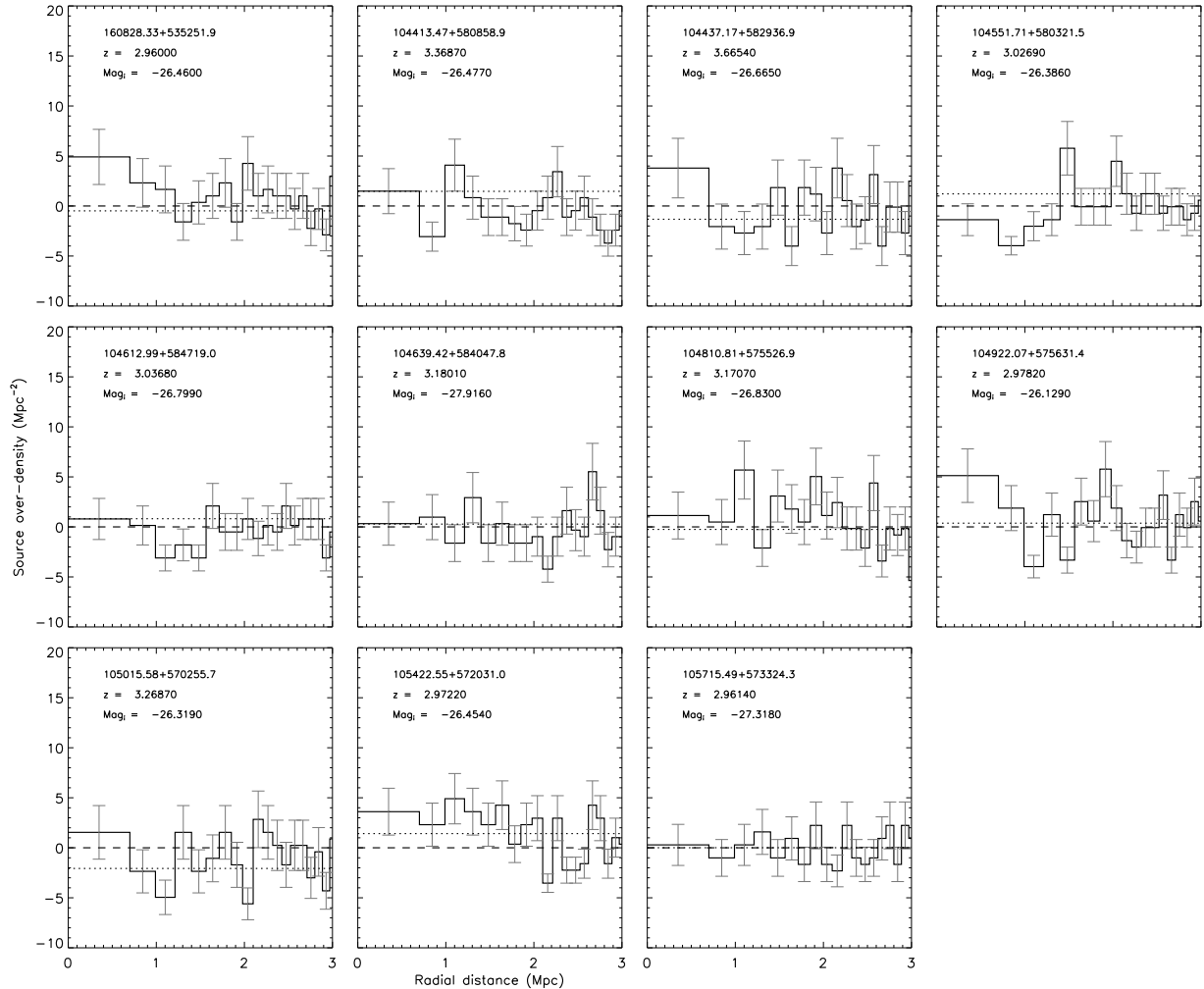}
\caption{Showing the individual source over-density vs radial distance for the
  11 QSOs in the redshift range $2.8 < z < 3.8$. The first bin has a radius of
  $700$~kpc and the other bins are of the same area as the first. The error
  bars show the Poisson error on the number counts. The dashed line shows the
  subtracted local background level (zero level) determined from an annulus of
  $2$~Mpc$-400$ arcsec from the QSOs. The dotted line shows, for comparison,
  the global background as determined by taking the average source density in
  large apertures over the SERVS fields. Also labelled are the QSO's redshifts
  and absolute SDSS $i$-band magnitudes. This is the source density before
  being corrected for completeness.}
\label{fig:individual_2_8-4_0}
\end{figure*}

\begin{figure}
\centering
\includegraphics[width=0.95\columnwidth]{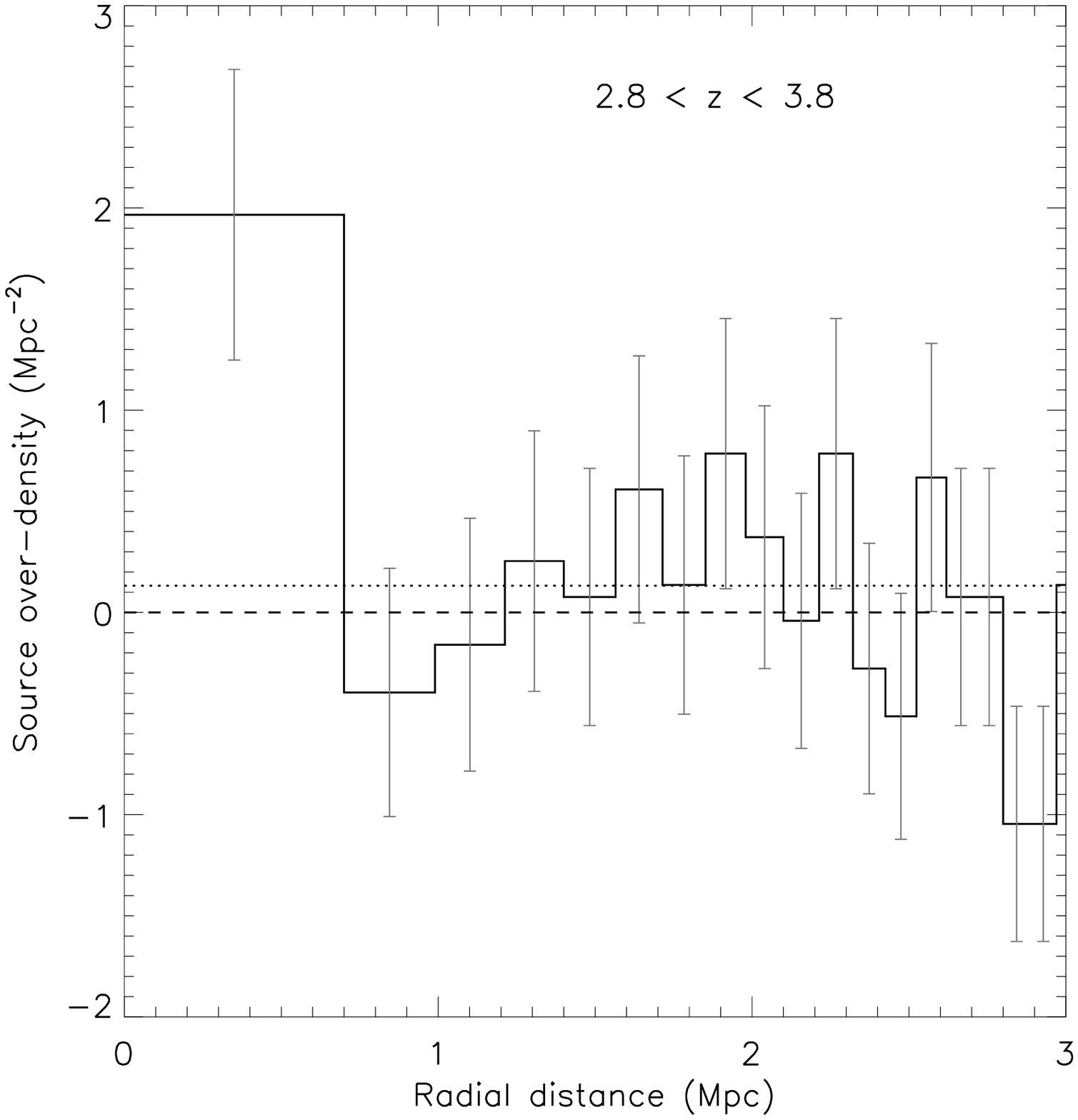}
\caption{Stacked source over-density vs radial distance for the 11 QSOs in the
  redshift range of $2.8 < z < 3.8$. The first bin has a radius of $700$~kpc
  and the other bins are of the same area as the first. The error bars show the
  Poisson error on the number counts. The dashed line shows the subtracted
  local background level (zero level) determined from an annulus of
  $2$~Mpc$-400$ arcsec from the QSOs. The dotted line shows, for comparison,
  the global background as determined from taking the average source density in
  large apertures over the SERVS fields. This is the source density before
  being corrected for completeness.}
\label{fig:2_8-4_0}
\end{figure}

\begin{figure}
\centering \includegraphics[width=0.95\columnwidth]{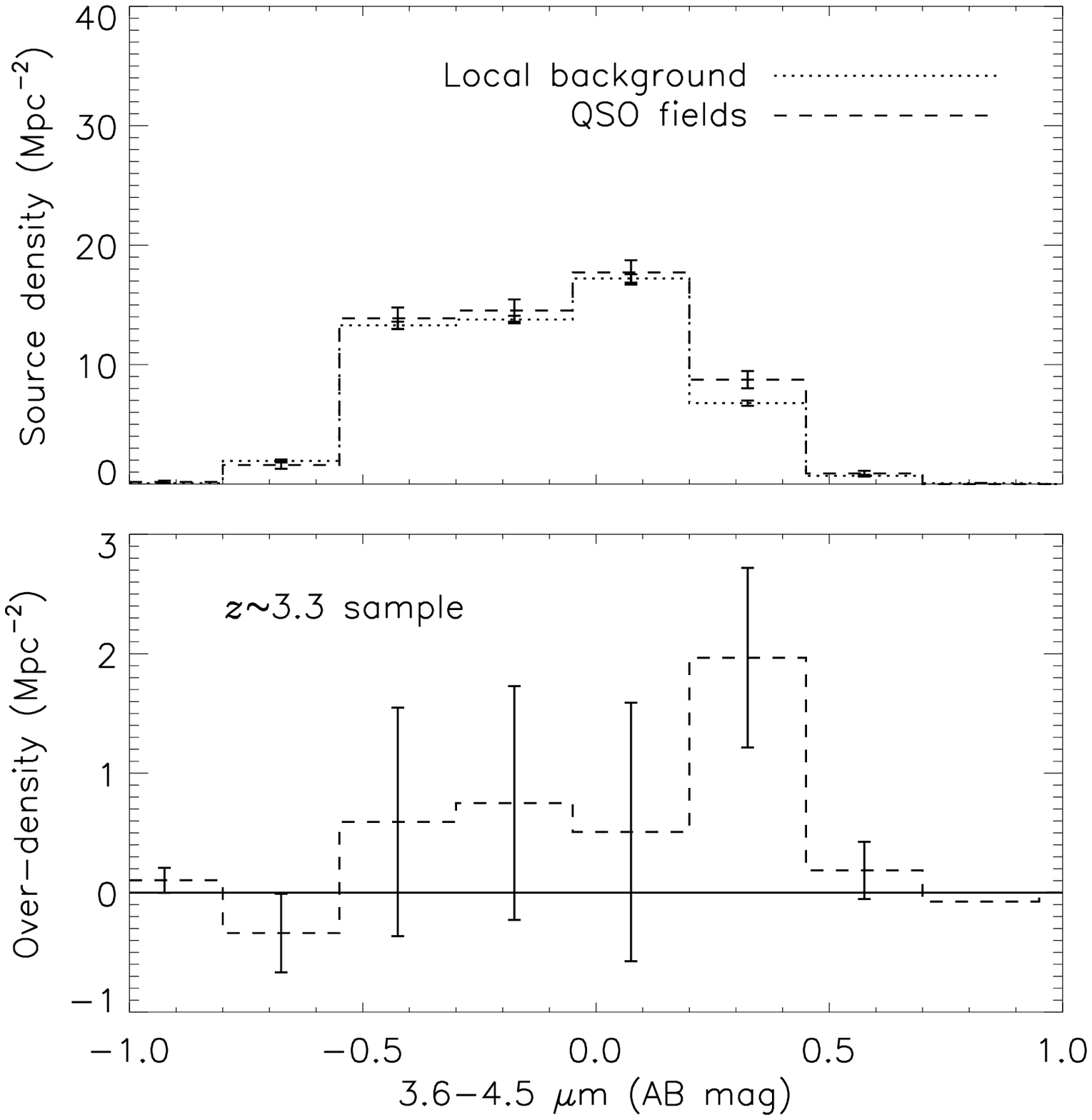}
\caption{Histograms showing the IRAC $3.6$-$4.5\mu$m colour of sources in the SERVS
  catalogues. In the top panel the dotted line shows the average local
  background source density surrounding the QSOs in the $z\sim3.3$ sample and
  the dashed line shows the average source density inside the first annulus
  surrounding these QSOs. The bottom panel shows the result of subtracting the
  local background from the source density first annulus.}
\label{fig:colour_histogram_z3}
\end{figure}

The individual histograms of source density versus radial distance for each of
these 11 QSOs are shown in the top panel of
Fig.~\ref{fig:individual_2_8-4_0}. Though there are no statistically
significant over-densities, many of the first annuli are more than 1-$\sigma$
(given by Poisson statistics) with one 2-$\sigma$ above the local background
level.

The stacked source density is plotted in Fig. \ref{fig:2_8-4_0} which shows a clear
peak in the source density within $700$~kpc of the QSOs. This peak is
significantly above the local background at the 2.62-$\sigma$ (given by Poisson
statistics). Looking at the individual histograms
(Fig.~\ref{fig:individual_2_8-4_0}) we again find that there is evidence to
suggest that using local background subtraction is the right approach. However,
using a global background in the stacking process has little affect on the
result, only reducing the significance to the 2.55-$\sigma$ level. We find that
if we allow our search to be slightly more sensitive by going down to the flux
at which we are 30 per cent complete the stacked source density in this colour
space becomes more significantly over-dense ($3.2$-$\sigma$), see Table
\ref{tab:midz}. This is further evidence that the result is likely to be real
since we get a stronger signal despite relaxing our conservative flux cut.
We use the source-density from the 50 per cent completeness
analysis for comparison to other work.

In Fig.~\ref{fig:colour_histogram_z3}, we show the $3.6$-$4.5\mu$m colour space
for the local backgrounds and first annulus for this sample along with the
result of subtracting the local background. There is a clear over-density at
the $\sim2.6$-$\sigma$ level in the colour range chosen. The result of this test is
thus consistent with the detected over-density being at the redshift of the
QSOs in this part of the sample.

When we run our Monte-Carlo code 1000 times on batches of 11 random locations,
avoiding these QSOs, we find that in this colour space we can generate similar
sized over-densities only 1.1 per cent of the time. If we extend this to all
the colour space used in the analysis (see Table \ref{tab:midz}) then this
increases to 11.9 per cent. Using the more sensitive 30 per cent completeness
flux limit this improves such that in the colour space chosen we only get a
$3.2$-$\sigma$ over-density 0.1 per cent of the time, and over all colour space
5 per cent of the time.

Physically this over-density with a correction for completeness corresponds to
on average 2-5 brighter than $\sim1.1L_*$ galaxies, with our choice of
models, in excess of the local field level around each QSO within
$\sim700$~kpc.

\subsection{Comparison between redshift bins}

To compare our two sub-samples we re-analyse the $z\sim2.0$ sample such that we
are only sensitive to galaxies with absolute magnitudes brighter than -24.4, to
match the sensitivity of the $z\sim3.3$ sample, again neglecting the
$k$-correction which is a 0.02 magnitude effect. In doing this we find 3-6
galaxies brighter $\sim L_*$ around the $z\sim2.0$ sample. Hence the population
of massive, brighter than $\sim L_*$, galaxies around the two samples seems to
be comparable. This is certainly good evidence that the massive (larger than
$L_*$) galaxies in proto-clusters are already in place by $z\sim3$-$4$, where
to date only a handful of detections have been made predominantly around
individual high-$z$ radio-galaxies (e.g.,
\citealt{Overzier06,Overzier08}). This picture fits in with the idea of
downsizing where massive galaxies form and cluster before those of lower mass
(\citealt{Cowie96};~\citealt{Heavens04}).

\section{Comparison with previous work at $\MakeLowercase{z}\sim1$.}
\label{z1}

In this section we compare our results to the previous work at $z\sim1$ by
\cite{Falder10}. The data for the $z\sim1$ sample is sensitive to galaxies with
an absolute magnitude of -23.0. We therefore looked again at the data from
\citet{Falder10} using a $700$ kpc annulus and restricting ourselves to flux
limits which mean that we our sampling the same absolute magnitude range at
$z\sim1$ as in the samples discussed here using the SERVS data. To compare with
the $z\sim2.0$ sample in which we are sensitive to an absolute magnitude of
-23.4 we apply a $k$-correction of 0.4 magnitudes and re-analyse the $z\sim1$
data down to an absolute magnitude of -23.0. This gives 4-6 galaxies within
$700$ kpc compared to the 7-12 around the $z\sim2.0$ sample. In the $z\sim3.3$
sample we are sensitive to galaxies with absolute magnitudes of -24.4, again
applying a $k$-correction of 0.4 magnitudes we re-analyse the $z\sim1$ data
down to an absolute magnitude of -24.0. This gives 1-2 galaxies within $700$
kpc compared to the 2-5 around the $z\sim3.3$ sample. On face value therefore
it suggests that the number of galaxies around $z\sim1$ sample is lower. This
may well be the case and would fit in with the idea of downsizing of the AGN
population where AGN at higher redshift are those in bigger groups or clusters
than at lower redshift \citep{Romano-Diaz11}. However it is likely that the
colour selection that we use in this work means we are more efficiently
removing contamination and therefore detecting a higher signal, this may go
some way to explaining the difference.

\section{Comparison with galaxy formation models}
\label{model}

In this section we compare our findings with predictions from the Durham
semi-analytic galaxy formation model of \citet{Bower06}, which is based on the
$\Lambda$CDM {\sc{millennium}} simulation \citep{Springel05}. They populate the
dark matter haloes created by the {\sc{millennium}} simulation with galaxies
using their semi-analytic formula, {\sc{galform}.} The {\sc{millennium}}
simulation is an N-body simulation consisting of a box with sides of
$500/h~$Mpc in co-moving units containing $10^{10}$ particles of mass $8.6
\times 10^8 {\Msolar}/h$. The simulation started from an initial set of density
perturbations at $z = 127$ calculated analytically, which is then allowed to
evolve under the influence of only gravity to the present day. Snapshot
catalogues of the structures (dark matter haloes) that formed and merged in the
box were saved at 64 epochs and it is these on to which the Durham galaxies are
added. The choice of the Durham model is partly based on the fact that they
give their galaxies central black-hole masses (see \citep{Bower06} for
details), which we use to compare to our QSOs, but also because it has recently
shown to be one of the best fitting models to the observed luminosity function
at $z<4$ \citep{Cir09}. However, as with all current models there are still
some issues with both the faint and bright end predictions
\citep[e.g.][]{Cir09,Henriques10}.

In order to compare our findings with what is predicted by the Durham model we
have queried the catalogues for each subset of our QSO sample. We have
generated catalogues from the model which contain objects with luminosities a
factor of $2$ fainter than we are sensitive to in each of the redshift
ranges. This allows us to show how the predictions change if we have under or
over-estimated our survey depth within a reasonable range. For each redshift
range we query the model catalogues at the closest snapshot to the mid-point of
the redshift range. We have then searched these catalogues for objects around
galaxies with black-hole masses greater than three different values ($M_{bh}
>$$10^{8.50}$, $10^{8.75}$ and $10^{9.0} \Msolar$).

This procedure should replicate as closely as possible what we have done with
the SDSS QSOs in SERVS, as by observing the fields around luminous high-$z$
QSOs we know we are centring on large black-holes at the redshift of the
QSO. It is not currently possible for most of our sample to measure the
black-hole masses using virial methods as even the \mg line moves out of the
SDSS spectral range at $z=2.25$. However, we know that these high luminosity
QSOs must be hosted by some of the largest black-holes at any given epoch. One
way to quantify this is through using Eddington arguments to place lower limits
on the black-hole masses of our QSOs. Using the SDSS absolute $i$-band
magnitude with a bolometric correction of 15 \citep{Richards06a} we can make
the assumption that the QSOs are accreting at the Eddington limit and so their
bolometric luminosity is equal to their Eddington luminosity. This then allows
us to place a lower limit on the mass of black-hole required to power each QSO,
using the relationship between Eddington luminosity and black-hole mass from
\citet{Rees84}. We present the results of this analysis on the right-hand axis
of Fig.~\ref{fig:L_z} showing the range of black-hole mass lower limits for
each part of the sample. It is worth noting that there are no black-holes in
the model at $z\sim3.3$ that have a mass $>10^{9.25}\Msolar$, hence why we
didn't extend the range of black holes we searched around to better match our
range in Fig.~\ref{fig:L_z}.

We are able to query the full simulation box for the higher redshift part of
our sample, but for the lower redshift part we have to restrict the volume
queried to a $350/h$~Mpc on a side box due to the time restriction on
queries. This is not an issue, however, because at lower redshift the
population of large black holes is, as would be expected, higher and so we
require less area to find the same number of suitable targets. As the longest
wavelength that the model produces is the $K$-band we use our modelled
$K-3.6{\mu}m$ colour from Section \ref{colour_cut} to match to our
observations. To account for the effect of in reality measuring counts in a
cylinder we search within our physical search radius of $700$~kpc in two
dimensions of the simulation and then within $3$~Mpc in the third
dimension. This effect as described in \citet{Yee&Ellingson95} increases the
sources detected by a factor of $\sim$1.5. We can then compare the measured
over-density, which we assume is all associated with the QSOs, to the number of
comparable galaxies we find in the model catalogues within the same radius.

\begin{figure*}
\centering
\includegraphics[width=0.95\columnwidth]{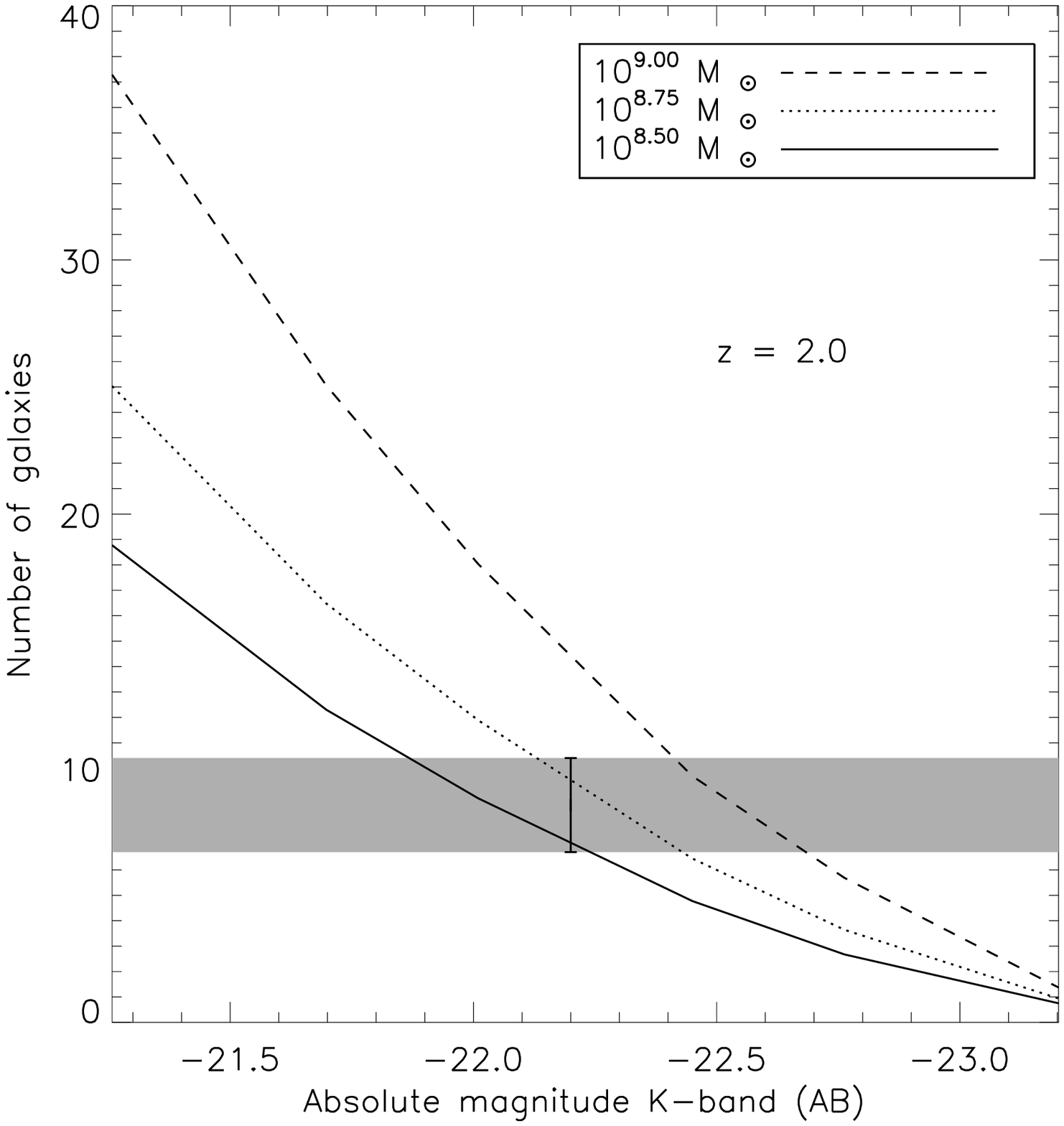}
\includegraphics[width=0.95\columnwidth]{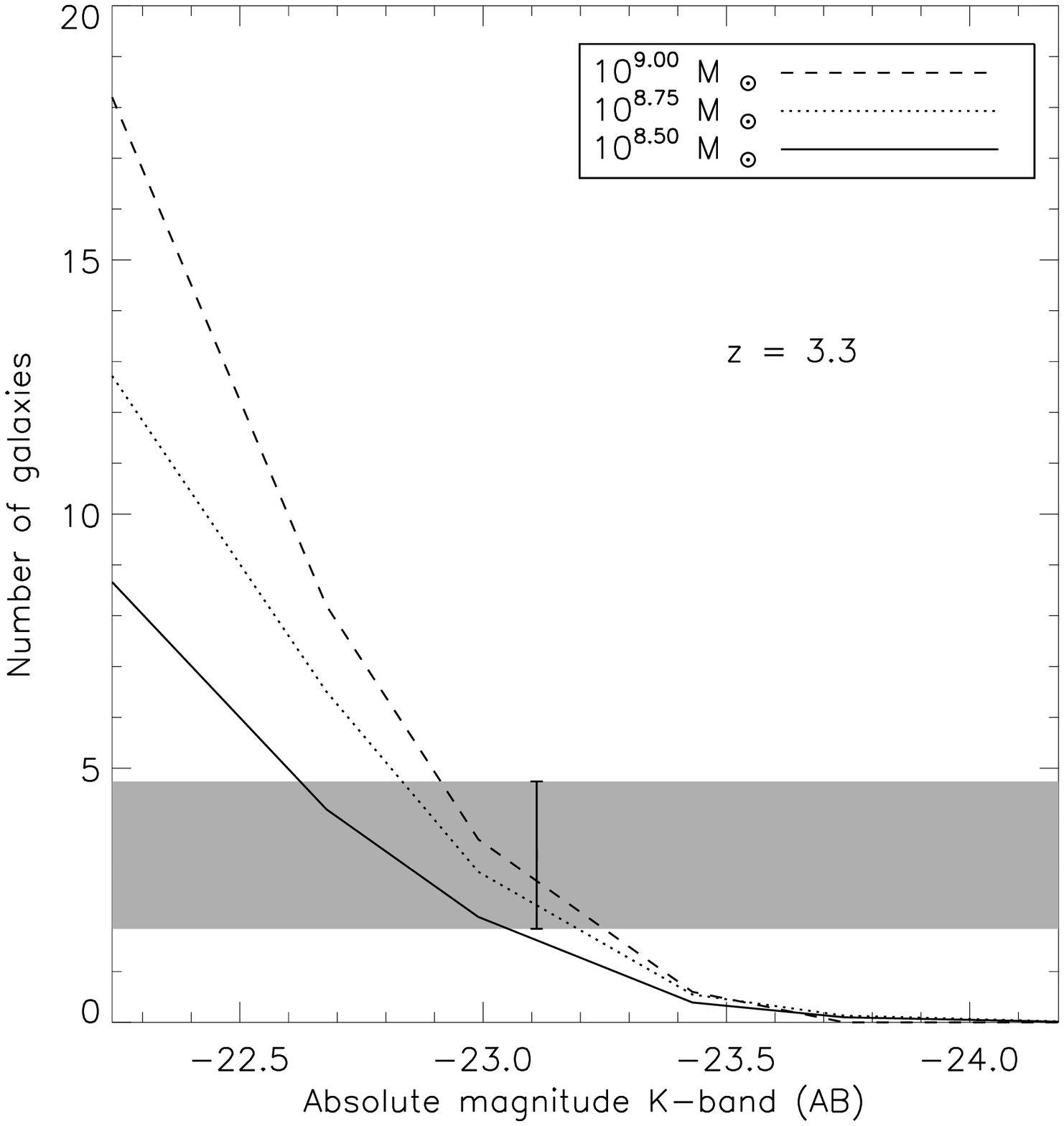}
\caption{The left panel shows the comparison of the $1.3 < z < 2.7$ sample to
  the Durham model, the right panel shows the same but for the $2.8 < z < 3.8$
  sample. The grey shaded band shows the 1-$\sigma$ error on the number of
  galaxies we have found surrounding our QSOs in each case, and the error bars
  show where we estimate we have reached down to on the galaxy luminosity
  function. These numbers are completeness corrected. The lines then show the
  number of galaxies predicted by the model within the same radius with three
  different masses for the central galaxy's black-hole mass (where the solid,
  dot, and dashed lines represent $10^{8.50}$, $10^{8.75}$ and $10^{9.0}
  \Msolar$ black-holes respectively).}
\label{fig:sims}
\end{figure*}

The results of this comparison are shown in Fig.~\ref{fig:sims}. The grey
shaded band shows the 1-$\sigma$ error on the number of galaxies we have found
surrounding our QSOs in each case, and the error bar shows where we estimate we
have reached down to in terms of galaxy absolute magnitude with our
analysis. The lines then show the number of galaxies predicted by the model
within the same cylindrical search area for galaxies with three different
central black-hole masses. 

Interestingly, in both cases we find that our detected source density matches
well with the predictions of the models. In the $z\sim2.0$ redshift bin
the model predictions for the $10^{8.50}$ and $10^{8.75}\Msolar$
black-holes fall within the $1$-$\sigma$ error bars of our measured source density. In
the $z\sim3.3$ redshift bin the model predictions for the $10^{8.75}$
and $10^{9.00}\Msolar$ black-holes fall within the $1$-$\sigma$ error bars of our
measured source density. It is worth noting that in our estimates of the
black-hole masses of our sample the $z\sim3.3$ part of the sample has on
average larger black-hole masses due to its on average higher luminosity.

\section{Summary}
\label{summary}

In this paper we have undertaken a study of the environments of SDSS QSOs in
the deep SERVS survey using data from {\em{Spitzer's}} IRAC instrument at
$3.6~{\mu}m$ and $4.5~{\mu}m$. We concentrate our study on the high-redshift
QSOs as these have not previously been studied with statistically large samples
or with data of this depth. These are highly luminous QSOs $M_i \lesssim -26$ and
hence harbour massive black-holes $M_{bh} \gtrsim 10^{8}\Msolar$. In contrast, the
environments of lower redshift QSOs have been studied in detail with much
larger samples \citep{Falder10}. We split the $z>1$
QSOs up into two sub-samples depending on their redshift, this allows us to
apply different source selection criteria to each sample. The criteria we apply
are a combination of an IRAC $3.6$-$4.5\mu$m colour selection and a cut of
sources detected above a certain brightness in the the ancillary $r$-band data
from the INT.

Using this method we are able to detect a significant ($>4$-$\sigma$)
over-density of galaxies around the QSOs in the sub-sample centred on
$z\sim2.0$ and ($>2$-$\sigma$) in the sub-sample centred on
$z\sim3.3$, providing furthur evidence that high luminosity AGN can be used to
trace clusters and proto clusters at these epochs. We compare the number counts
of $L_*$ or brighter galaxies around each sample and find them to be
comparable, suggesting the massive galaxies in proto-clusters are in place by
$z\sim3-4$ which is consistent with the idea of downsizing. We then compare
these findings to those of \citet{Falder10} at $z\sim1$ and find that the
over-densities found in this work are slightly larger than those found in
\citet{Falder10}. However it is likely that the colour selection used in this
work allows for a more efficient detection of possible companion galaxies and
therefore this difference may well not be a real effect.

We then compare our results to the predictions from the Durham \citep{Bower06}
galaxy formation model, built on top of the {\sc{Millennium}} simulation
\citep{Springel05} dark matter halo catalogues. In both cases we find the model
predictions are within the $1$-$\sigma$ error bars of our measured source
density.

\acknowledgments

We thank the anonymous referee for comments that improved the paper. This
work is based [in part] on observations made with the {\em Spitzer Space Telescope} 
which is operated by the Jet Propulsion Laboratory, California
Institute of Technology under a contract with NASA. Support for this work was
provided by NASA through an award issued by JPL/Caltech. JTF thanks the Science
and Technology Facilities Council for a research studentship. JA gratefully
acknowledges the support from the Science and Technology Foundation (FCT,
Portugal) through the research grant PTDC/FIS/100170/2008.

\bibliography{../bibtex}{} \bibliographystyle{apj}

\end{document}